\documentclass[aps,prl,twocolumn,showpacs,showkeys]{revtex4-1}
\usepackage{mathtools}

\begin{document}

%\preprint{}

\title{Another Formulation of the Wick's Theorem. Farewell, Pairing?}


\author{Igor~V.~Beloussov}
\email[]{igor.beloussov@phys.asm.md}
%\thanks{}
\affiliation{Institute of Applied Physics, Academy of Sciences of Moldova,\\
5 Academy Str., Kishinev, 2028, Republic of Moldova}


\date{\today}

\begin{abstract}
The algebraic formulation of Wick's theorem that allows one to present the vacuum or thermal averages of the chronological product of an arbitrary number of field operators as a determinant (permanent) of the matrix is proposed. Each element of the matrix is the average of the chronological product of only two operators. This formulation is extremely convenient for practical calculations in quantum field theory and statistical physics by the methods of symbolic mathematics using computers.
\end{abstract}

%\keywords{}

\maketitle

\section{\label{Int}Introduction}

Wick's theorems are used extensively in quantum field theory~\cite{Weinberg1995,Schweber1961,Peskin1995,Bogoliubov1980} and statistical physics~\cite{Negele1998,A.A.Abrikosov1963,Fetter1971}. They allow one to use the Green's functions method, and consequently to apply the Feynman's diagrams for investigations~\cite{Weinberg1995,Schweber1961,Peskin1995}. The first of these, which can be called {\it Wick's Theorem for Ordinary Products}\/, gives us the opportunity to reduce in almost automatic mode the usual product of operators into a unique sum of normal products multiplied by  $c$--numbers. It can be formulated as follows~\cite{Bogoliubov1980}.
Let   ${{A}_{i}}\left( {{x}_{i}} \right)$  ($i=1,2,\ldots ,n$ )
are ``linear operators'', i.e., some linear combinations of creation and annihilation operators. Then {\it the ordinary product of linear operators is equal to the sum of all the corresponding normal products with all possible contractions, including the normal product without contractions, i.e.,}\/
\begin{eqnarray*}
A_{1}\ldots A_{n} &=&\colon A_{1}\ldots A_{n}\colon +\colon %
\underbracket[0.5pt]{A_{1}A_{2}}\ldots A_{n}\colon +\ldots  \\
&&+\colon \underbracket[0.5pt]{A_{1}\ldots A_{n-1}}A_{n}\colon +\colon %
\underbracket[0.5pt]{A_{1}\ldots A_{n}}\colon  \\
&&+\colon \underbracket[0.5pt]{A_{1}A_{2}}\underbracket[0.5pt]{A_{3}A_{4}}\ldots
A_{n}\colon +\ldots \;,
\end{eqnarray*}
where
$\underbracket[0.5pt]{A_{i}A_{j}}=A_{i}A_{j}-\colon A_{i}A_{j}\colon$
($ i,j=1,2,\ldots ,n$)
is the contraction between the factors $A_{i}$ and $A_{j}$. Since the vacuum expectation value of the normal ordered product is equal zero, this theorem provides us a way of expressing the vacuum expectation values of $n$  linear operators in terms of vacuum expectation values of two operators.

{\it Wick's Theorem for Chronological Products}\/~\cite{Bogoliubov1980} asserts that {\it the  $T$--product of a system of $n$  linear operators is equal to the sum of their normal products with all possible chronological contractions, including the term without contractions}\/. It follows directly from the previous theorem and gives the opportunity to calculate the vacuum expectation values of the chronological products of linear operators.

Finally, from Wick's theorem for chronological products the {\it Generalized Wick's Theorem}\/~\cite{Bogoliubov1980} can be obtained.  It asserts that  {\it the vacuum expectation value of the chronological product of $n+1$  linear operators $A,{{B}_{1}},\ldots ,{{B}_{n}}$  can be decomposed into the sum of $n$  vacuum expectation values of the same chronological products with all possible contractions of one of these operators (for example $A$ ) with all others, i.e.}\/,
\begin{equation}
{{\left\langle T\left( A{{B}_{1}}\ldots {{B}_{n}}\right) \right\rangle }_{0}}%
=\sum\limits_{1\leq i\leq n}{{{\left\langle T\left( \overbracket[0.5pt]{
AB_{1}\ldots B_{i}}\ldots {{B}_{n}}\right) \right\rangle }_{0}}}\,\,.
\label{gen}
\end{equation}
Here
$
\overbracket[0.5pt]{ A_{i} A_{i}}=T\left( {A}_{i}{A}_{j}\right) -:{A}_{i}{A}_{j}:={{%
\left\langle T\left( {A}_{i}{A}_{j}\right) \right\rangle }_{0}}
$  ($ i,j=1,2,\ldots ,n$ ) is the chronological contraction between the factors  $A_{i}$ and $A_{j}$.
It should be noted that, in contrast to the usual Wick's theorem for chronological products, there are no expressions involving a number of contractions greater than one on the right-hand side of (\ref{gen}).

Wick's theorem for chronological products or its ge\-ne\-ra\-lized version are used for the calculation of matrix elements of the scattering matrix in each order of perturbation theory~\cite{Weinberg1995,Schweber1961,Peskin1995,Bogoliubov1980}. The procedure is reduced to calculation of the vacuum expectation of chronological products of the field operators in the interaction representation. As factors in these products a number of operators $\psi _i$  of the Fermi fields and the same number of their “conjugate” operators  $\bar\psi _i$ , as well as operators of the Bose fields  ${{\varphi }_{s}}=\varphi _{s}^{\left( + \right)}+\varphi _{s}^{\left( -
\right)}$ may be used. Here all continuous and discrete variables are included in the index. In the interaction representation the operators ${{\psi }_{i}}$,${{\bar{\psi}}_{i}}$, and ${{\varphi }_{s}^{\left( \pm
\right) }}$ correspond to free fields and satisfy the commutation relationships of the form
${{\left[ {{\psi }_{i}},\,{{\psi }_{j}} \right]}_{+}}={{\left[ {{{\bar{\psi }%
}}_{i}},\,{{{\bar{\psi }}}_{j}} \right]}_{+}}=0$, ${{\left[ {{\psi }_{i}},\,{%
{{\bar{\psi }}}_{j}} \right]}_{+}}={{D}_{ij}}$, ${{\left[ \varphi
_{r}^{\left( - \right)},\,\varphi _{s}^{(-)} \right]}_{-}}={{\left[ \varphi
_{r}^{\left( + \right)},\,\varphi _{s}^{(+)} \right]}_{-}}=0$, ${{\left[
\varphi _{r}^{\left( + \right)},\,\varphi _{s}^{(-)} \right]}_{-}}={{\bar{D}}%
_{rs}}$. Therefore, the averaging of the Fermi and Bose fields can be performed independently.

\section{\label{Theorem}Basic theorem}

Since we may rearrange the order of the operators inside $T$--products taking into account the change of the sign, which arises when the order of the Fermi operators is changed, we present our vacuum expectation value of the chronological product of the Fermi operators in the form
\begin{equation}
\pm {{\left\langle T\left[ \left( {{\psi }_{{{i}_{1}}}}{{{\bar{\psi}}}_{{{j}%
_{1}}}}\right) \left( {{\psi }_{{{i}_{2}}}}{{{\bar{\psi}}}_{{{j}_{2}}}}%
\right) \ldots \left( {{\psi }_{{{i}_{n}}}}{{{\bar{\psi}}}_{{{j}_{n}}}}%
\right) \right] \right\rangle }_{0}}\,.  \label{averge}
\end{equation}
To calculate (\ref{averge}), we can use Wick's theorem for chronological products. However, while considering the higher-order perturbation theory, the number of pairs $\psi _i \bar\psi _j$ of operators $\psi _i$  and $\bar\psi _j$  becomes so large that the direct application of this theorem begins to represent certain problems because it is very difficult to sort through all the possible contractions between  $\psi _i$  and $\bar\psi _j$.

	A consistent use of generalized Wick's theorem would introduce a greater accuracy in our actions. However, in this case we expect very cumbersome and tedious calculations. Hereinafter we show that the computation of (\ref{averge}) can be easily performed using a simple formula
\begin{equation}
{{\left\langle T\left[ \left( {{\psi }_{{{i}_{1}}}}{{{\bar{\psi}}}_{{{j}_{1}}%
}}\right)
\ldots \left( {{\psi }_{{{i}_{n}}}}{{{\bar{\psi}}}_{{{j}_{n}}}}\right) %
\right] \right\rangle }_{0}}=\det \left( {{\Delta }_{{{i}_{\alpha }}{{j}%
_{\beta }}}}\right) \ ,  \label{fermi}
\end{equation}
where
\begin{eqnarray}
{\Delta }_{{i}_{\alpha }{{j}_{\beta }}} &=&\overbracket[0.5pt]{{\psi }_{{i}_{\alpha
}} {\bar{\psi}}_{{j}_{\beta }}}={\left\langle T\left( {\psi }_{{i}_{\alpha }}%
{\bar{\psi}}_{{j}_{\beta }}\right) \right\rangle }_{0}  \nonumber \\
&&\left( \alpha ,\beta =1,2,\ldots ,n\right) \ \ .  \label{contr}
\end{eqnarray}
The proof of this theorem is by induction. Let us assume now that (\ref{fermi}) is true for $n$  pairs ${\psi }_{i_\alpha }
{\bar{\psi}}_{j_\beta}$ , and consider it for the case  $n+1$. Using generalized Wick's theorem we have
\begin{widetext}
\begin{gather*}
{{\left\langle T\left[ \left( {{\psi }_{{{i}_{1}}}}{{{\bar{\psi}}}_{{{j}_{1}}%
}}\right) \left( {{\psi }_{{{i}_{2}}}}{{{\bar{\psi}}}_{{{j}_{2}}}}\right)
\ldots \left( {{\psi }_{{{i}_{n\ast 1}}}}{{{\bar{\psi}}}_{{{j}_{n\ast 1}}}}%
\right) \right] \right\rangle }_{0}}=-{\Delta }_{{i}_{1}{{j}_{n+1}}}{{%
\left\langle T\left[ \left( {{\psi }_{{{i}_{n\ast 1}}}}{{{\bar{\psi}}}_{{{j}%
_{1}}}}\right) \left( {{\psi }_{{{i}_{2}}}}{{{\bar{\psi}}}_{{{j}_{2}}}}%
\right) \ldots \left( {{\psi }_{{{i}_{n}}}}{{{\bar{\psi}}}_{{{j}_{n}}}}%
\right) \right] \right\rangle }_{0}} \\
-\sum\limits_{\gamma =2}^{n-1}{\Delta }_{{i}_{\gamma }{{j}_{n+1}}}{{%
\left\langle T\left[ \left( {{\psi }_{{{i}_{1}}}}{{{\bar{\psi}}}_{{{j}_{1}}}}%
\right) \ldots \left( {{\psi }_{{{i}_{\gamma -1}}}}{{{\bar{\psi}}}_{{{j}%
_{\gamma -1}}}}\right) \left( {{\psi }_{{{i}_{n+1}}}}{{{\bar{\psi}}}_{{{j}%
_{\gamma }}}}\right) \left( {{\psi }_{{{i}_{\gamma +1}}}}{{{\bar{\psi}}}_{{{j%
}_{\gamma +1}}}}\right) \ldots \left( {{\psi }_{{{i}_{n}}}}{{{\bar{\psi}}}_{{%
{j}_{n}}}}\right) \right] \right\rangle }}_{0} \\
-{\Delta }_{{i}_{n}{{j}_{n+1}}}{{\left\langle T\left[ \left( {{\psi }_{{{i}%
_{1}}}}{{{\bar{\psi}}}_{{{j}_{1}}}}\right) \ldots \left( {{\psi }_{{{i}_{n-1}%
}}}{{{\bar{\psi}}}_{{{j}_{n-1}}}}\right) \left( {{\psi }_{{{i}_{n+1}}}}{{{%
\bar{\psi}}}_{{{j}_{n}}}}\right) \right] \right\rangle }}_{0}+{\Delta }_{{i}%
_{n+1}{{j}_{n+1}}}{{\left\langle T\left[ \left( {{\psi }_{{{i}_{1}}}}{{{\bar{%
\psi}}}_{{{j}_{1}}}}\right) \ldots \left( {{\psi }_{{{i}_{n}}}}{{{\bar{\psi}}%
}_{{{j}_{n}}}}\right) \right] \right\rangle }}_{0}\;.
\end{gather*}
\end{widetext}
Taking into account (\ref{fermi}) we obtain
\begin{widetext}
\begin{gather}
{{\left\langle T\left[ \left( {{\psi }_{{{i}_{1}}}}{{{\bar{\psi}}}_{{{j}_{1}}%
}}\right) \left( {{\psi }_{{{i}_{2}}}}{{{\bar{\psi}}}_{{{j}_{2}}}}\right)
\ldots \left( {{\psi }_{{{i}_{n\ast 1}}}}{{{\bar{\psi}}}_{{{j}_{n\ast 1}}}}%
\right) \right] \right\rangle }_{0}}=  \nonumber \\
-{\Delta }_{{i}_{1}{{j}_{n+1}}}\left\vert
\begin{array}{llll}
{\Delta }_{{i}_{n+1}{{j}_{1}}} & {\Delta }_{{i}_{n+1}{{j}_{2}}} & \cdots  & {%
\Delta }_{{i}_{n+1}{{j}_{n}}} \\
{\Delta }_{{i}_{2}{{j}_{1}}} & {\Delta }_{{i}_{2}{{j}_{2}}} & \cdots  & {%
\Delta }_{{i}_{2}{{j}_{n}}} \\
\;\vdots  & \;\vdots  & \ddots  & \;\vdots  \\
{\Delta }_{{i}_{n}{{j}_{1}}} & {\Delta }_{{i}_{n}{{j}_{2}}} & \cdots  & {%
\Delta }_{{i}_{n}{{j}_{n}}}%
\end{array}%
\right\vert -{\Delta }_{{i}_{2}{{j}_{n+1}}}\left\vert
\begin{array}{llll}
{\Delta }_{{i}_{1}{{j}_{1}}} & {\Delta }_{{i}_{1}{{j}_{2}}} & \cdots  & {%
\Delta }_{{i}_{1}{{j}_{n}}} \\
{\Delta }_{{i}_{n+1}{{j}_{1}}} & {\Delta }_{{i}_{n+1}{{j}_{2}}} & \cdots  & {%
\Delta }_{{i}_{n+1}{{j}_{n}}} \\
\;\vdots  & \;\vdots  & \ddots  & \;\vdots  \\
{\Delta }_{{i}_{n}{{j}_{1}}} & {\Delta }_{{i}_{n}{{j}_{2}}} & \cdots  & {%
\Delta }_{{i}_{n}{{j}_{n}}}%
\end{array}%
\right\vert -\ldots   \label{exp2} \\
-{\Delta }_{{i}_{n}{{j}_{n+1}}}\left\vert
\begin{array}{llll}
{\Delta }_{{i}_{1}{{j}_{1}}} & {\Delta }_{{i}_{1}{{j}_{2}}} & \cdots  & {%
\Delta }_{{i}_{1}{{j}_{n}}} \\
\;\vdots  & \;\vdots  & \ddots  & \;\vdots  \\
{\Delta }_{{i}_{n-1}{{j}_{1}}} & {\Delta }_{{i}_{n-1}{{j}_{2}}} & \cdots  & {%
\Delta }_{{i}_{n-1}{{j}_{n}}} \\
{\Delta }_{{i}_{n+1}{{j}_{1}}} & {\Delta }_{{i}_{n+1}{{j}_{2}}} & \cdots  & {%
\Delta }_{{i}_{n+1}{{j}_{n}}}%
\end{array}%
\right\vert +{\Delta }_{{i}_{n+1}{{j}_{n+1}}}\left\vert
\begin{array}{llll}
{\Delta }_{{i}_{1}{{j}_{1}}} & {\Delta }_{{i}_{1}{{j}_{2}}} & \cdots  & {%
\Delta }_{{i}_{1}{{j}_{n}}} \\
{\Delta }_{{i}_{2}{{j}_{1}}} & {\Delta }_{{i}_{2}{{j}_{2}}} & \cdots  & {%
\Delta }_{{i}_{2}{{j}_{n}}} \\
\;\vdots  & \;\vdots  & \ddots  & \;\vdots  \\
{\Delta }_{{i}_{n}{{j}_{1}}} & {\Delta }_{{i}_{n}{{j}_{2}}} & \cdots  & {%
\Delta }_{{i}_{n}{{j}_{n}}}%
\end{array}%
\right\vert \;.  \nonumber
\end{gather}
\end{widetext}
Rearranging the rows in the determinants in (\ref{exp2}) it is easy to see that the right hand side is the expansion of the
\begin{gather*}
\det \left( {{\Delta }_{{{i}_{\alpha }}{{j}_{\beta }}}}\right) =\left\vert
\begin{array}{llll}
{\Delta }_{{i}_{1}{{j}_{1}}} & {\Delta }_{{i}_{1}{{j}_{2}}} & \cdots  & {%
\Delta }_{{i}_{1}{{j}_{n+1}}} \\
{\Delta }_{{i}_{2}{{j}_{1}}} & {\Delta }_{{i}_{2}{{j}_{2}}} & \cdots  & {%
\Delta }_{{i}_{2}{{j}_{n+1}}} \\
\;\vdots  & \;\vdots  & \ddots  & \;\vdots  \\
{\Delta }_{{i}_{n+1}{{j}_{1}}} & {\Delta }_{{i}_{n+1}{{j}_{2}}} & \cdots  & {%
\Delta }_{{i}_{n+1}{{j}_{n+1}}}%
\end{array}%
\right\vert  \\
\left( \alpha ,\beta =1,2,\ldots ,n+1\right)
\end{gather*}
along the last column~\cite{Korn1961}. The validity of (\ref{fermi}) for $n=1$  follows from the definition
$\Delta _{
i_{\alpha} j_{\beta }
}$
in (\ref{contr}).

Note that this result does not depend on the way how we divide the operators on the left hand side of  (\ref{fermi}) into pairs $\psi_{i_\alpha} \bar\psi_{j_\beta}$. Indeed, if on the left side of (\ref{fermi}) we permute, for example, $\bar\psi_{j_\beta}$  and $\bar\psi_{j_\gamma}$  ($\beta \neq \gamma$), it changes its sign. The same happens on the right hand side of (\ref{fermi}) since this change leads to the permutation of two columns in the determinant, and it also changes its sign. Similarly, in the case of a permutation  of $\psi_{i_\alpha} $ and $\psi_{i_\delta} $  ($\alpha \neq \delta$). Obviously, when the whole pair $\psi_{i_\alpha} \bar\psi_{j_\beta}$  is transposed, the left and right hand sides of (\ref{fermi}) do not change.\\

Obviously, the formula similar to (3) can be obtained and in the case of Bose fields
\begin{gather}
{{\left\langle T\left[ \left( {{\varphi }_{{{i}_{1}}}^{\left( +\right) }{%
\varphi }_{{{j}_{1}}}^{\left( -\right) }}\right) \left( {{\varphi }_{{{i}_{2}%
}}^{\left( +\right) }{\varphi }_{{{j}_{2}}}^{\left( -\right) }}\right)
\ldots \left( {{\varphi }_{{{i}_{n}}}^{\left( +\right) }{\varphi }_{{{j}_{n}}%
}^{\left( -\right) }}\right) \right] \right\rangle }_{0}} \nonumber \\
=\mathrm{perm}\left( {{\Delta }_{{{i}_{\alpha }}{{j}_{\beta }}}}\right) \ , \label{perm}
\end{gather}
where
\begin{gather}
{\bar{\Delta}}_{{i}_{\alpha }{{j}_{\beta }}}=\overbracket[0.5pt]{{\varphi}^{\left(
+ \right)}_{{i}_{\alpha }} {{\varphi}}^{\left( - \right)}_{{j}_{\beta }}}={%
\left\langle T\left( {{\varphi }_{{{i}_{\alpha }}}^{\left( +\right) }{%
\varphi }_{{{j}_{\beta }}}^{\left( -\right) }}\right) \right\rangle }_{0}
\nonumber \\
\left( \alpha ,\beta =1,2,\ldots ,n\right) \ \ .
\end{gather}

In quantum statistics the  $n$--body thermal, or imaginary-time, Green's functions in the {\it Grand Canonical Ensemble}\/ are defined as the thermal trace of a time-ordered product of the fields operators in the imaginary-time Heisenberg representation~\cite{Negele1998,A.A.Abrikosov1963,Fetter1971}. To calculate them in each order of perturbation theory, Wick's theorem is also used. Obviously, in this case, the theorem also may be formulated in the form (\ref{fermi}) and (\ref{perm}) convenient for practical calculation.

\section{\label{exam}Simple examples}

In order to demonstrate the usability of the proposed formulation of Wick's theorem, we find the first-order correction to the one- and two-particle thermal Green’s functions for the Fermi system described in the interaction representation by the Hamiltonian
\begin{eqnarray*}
{{H}_{int}}\left( \tau \right)  &=&\frac{1}{2}\int {d{{\mathbf{r}}_{1}}d{{%
\mathbf{r}}_{2}}\,{{{\bar{\psi}}}_{\alpha }}\left( {{\mathbf{r}}_{1}},\tau
\right) }{{\bar{\psi}}_{\beta }}\left( {{\mathbf{r}}_{2}},\tau \right)  \\
&&\times U\left( {{\mathbf{r}}_{1}}-{{\mathbf{r}}_{2}}\right) {{\psi }%
_{\beta }}\left( {{\mathbf{r}}_{2}},\tau \right) {{\psi }_{\alpha }}\left( {{%
\mathbf{r}}_{1}},\tau \right)
\end{eqnarray*}
that contains the product of the field operators ${{\psi }_{\alpha }}\left( {{\mathbf{r}}_{1}},\tau \right)$ and ${{\bar{\psi
}}_{\alpha }}\left( {{\mathbf{r}}_{1}},\tau \right)$   in this representation (parameter $\alpha$  indicates the spin projections, $\tau$  is the imaginary-time). The one-particle Green's function can be represented as~\cite{Negele1998,A.A.Abrikosov1963}
\[
\mathcal{G}_{I}\left( {{x}_{1}},{{x}_{2}} \right)= -\frac{{{\left\langle {{T}%
_{\tau }}\left[ \psi \left( {{x}_{1}} \right)\bar{\psi }\left( {{x}_{2}}
\right)\mathcal{S} \right] \right\rangle }_{0}}}{{{\left\langle \mathcal{S}
\right\rangle }_{0}}}\,,\,
\]
where ${{\left\langle \ldots \right\rangle }_{0}}$ is the symbol for the Gibbs average over the states of a system of noninteracting particles, $x\equiv \left( \mathbf{r},\tau ,\alpha \right)$ and
\[
\mathcal{S}\left( \tau \right) ={{T}_{\tau }}\exp \left\{
-\int\limits_{0}^{\tau }{d\tau ^{\prime }\,{{H}_{int}}\left( \tau
^{\prime }\right) \,}\right\} \,.
\]
We obtain
\begin{gather}
\mathcal{G}_{I}\left( {{x}_{1}},{{x}_{2}}\right) =-{{\left\langle {T}_{\tau }%
\left[ {{\psi \left( {{x}_{1}}\right) \bar{\psi}}}\left( {{x}_{2}}\right) %
\right] \right\rangle }_{0}}  \nonumber \\
+\frac{1}{2{{\left\langle \mathcal{S}\right\rangle }_{0}}}\int
dz_{1}dz_{2}\mathcal{V}\left( z_{1}-z_{2}\right)   \nonumber \\
\times {{\left\langle {T}_{\tau }\left[ \left( {{\psi \left( {{x}_{1}}%
\right) \bar{\psi}}}\left( {{x}_{2}}\right) \right) \left( {{\psi \left( {%
z_{1}}\right) {\bar{\psi}}\left( {z_{1}}\right) }}\right) \left( {{\psi
\left( {z}_{2}\right) {\bar{\psi}}}}\left( z_{2}\right) \right) \right]
\right\rangle }_{0}}  \nonumber \\
=-\Delta \left( {{x}_{1}},{{x}_{2}}\right) +\frac{1}{2{{\left\langle
\mathcal{S}\right\rangle }_{0}}}\int dz_{1}dz_{2}\mathcal{V}\left( z_{1}-z_{2}\right)
\nonumber \\
\times
\begin{vmatrix}
\Delta \left( {{x}_{1}},{{x}_{2}}\right)  & \Delta \left( {{x}_{1}},{{z}_{1}}%
\right)  & \Delta \left( {{x}_{1}},{z_{2}}\right)  \\
\Delta \left( {z_{1}},{{x}_{2}}\right)  & \Delta \left( {z_{1}},{{z}_{1}}%
\right)  & \Delta \left( {z_{1}},{z_{2}}\right)  \\
\Delta \left( {z_{2}},{{x}_{2}}\right)  & \Delta \left( {z_{2}},{{z}_{1}}%
\right)  & \Delta \left( {z_{2}},{z_{2}}\right)
\end{vmatrix}%
\;,  \label{ggg}
\end{gather}
where $\mathcal{V}\left( {{x}_{1}}-{{x}_{2}} \right)=U\left( {{\mathbf{r}}_{1}}-{{%
\mathbf{r}}_{2}} \right)\delta \left( {{\tau }_{1}}-{{\tau }_{2}} \right)$.
We can immediately take into account the reduction of the disconnected diagrams, if we assume in (\ref{ggg})  ${\left\langle \mathcal{S}
\right\rangle }_{0}=1$ and $\Delta \left( {{x}_{1}},{{x}_{2}}\right)=0$~\cite{A.A.Abrikosov1963}. Then,
\begin{gather*}
\mathcal{G}_{I}\left( {{x}_{1}},{{x}_{2}}\right) =-\Delta \left( {{x}_{1}},{{%
x}_{2}}\right) +\frac{1}{2{{\left\langle \mathcal{S}\right\rangle }_{0}}}%
\int dz_{1}dz_{2}\mathcal{V}\left( z_{1}-z_{2}\right)  \\
\times\left[ \Delta \left( {{x}_{1}},{z_{2}}\right)
\begin{vmatrix}
\Delta \left( {z_{1}},{{x}_{2}}\right)  & \Delta \left( {z_{1}},{{z}_{1}}%
\right)  \\
\Delta \left( {z_{2}},{{x}_{2}}\right)  & \Delta \left( {z_{2}},{{z}_{1}}%
\right)
\end{vmatrix}%
\right.  \\
\left. -\Delta \left( {{x}_{1}},{{z}_{1}}\right)
\begin{vmatrix}
\Delta \left( {z_{1}},{{x}_{2}}\right)  & \Delta \left( {z_{1}},{z_{2}}%
\right)  \\
\Delta \left( {z_{2}},{{x}_{2}}\right)  & \Delta \left( {z_{2}},{z_{2}}%
\right)
\end{vmatrix}%
\right]  \\
=-\Delta \left( {{x}_{1}},{{x}_{2}}\right)  \\
+\int dz_{1}dz_{2}\Delta \left( {{x}_{1}},{{z}_{1}}\right) \mathcal{V}\left(
z_{1}-z_{2}\right) \Delta \left( {z_{1}},{z_{2}}\right) \Delta \left( {z_{2}}%
,{{x}_{2}}\right)  \\
-\int dz_{1}dz_{2}\Delta \left( {{x}_{1}},{{z}_{1}}\right) \mathcal{V}\left(
z_{1}-z_{2}\right) \Delta \left( {z_{2}},{z_{2}}\right) \Delta \left( {z_{1}}%
,{{x}_{2}}\right) \;.
\end{gather*}%
Taking into account  ${{\mathcal{G}}^{\left( 0\right) }}\left( {{x}_{1}},%
{{x}_{2}}\right) =-\Delta \left( {{x}_{1}},{{x}_{2}}\right) $  we obtain
finally
\begin{gather*}
\mathcal{G}_{I}\left( {{x}_{1}},{{x}_{2}}\right) ={{\mathcal{G}}^{\left(
0\right) }}\left( {{x}_{1}},{{x}_{2}}\right)  \\
+\int dz_{1}dz_{2}{{\mathcal{G}}^{\left( 0\right) }}\left( {{x}_{1}},{z_{1}}%
\right) \Sigma ^{\left( 1\right) }\left( {z_{1}},{z_{2}}\right) _{2}{{%
\mathcal{G}}^{\left( 0\right) }}\left( {z_{2}},x{_{2}}\right) \;,
\end{gather*}%
\begin{gather*}
\Sigma ^{\left( 1\right) }\left( {z_{1}},{z_{2}}\right) =-\mathcal{V}\left(
z_{1}-z_{2}\right) {{\mathcal{G}}^{\left( 0\right) }}\left( {z_{1}},{z_{2}}%
\right)  \\
+\delta \left( z_{1}-z_{2}\right) \int dz\mathcal{V}\left( z_{1}-z\right) {{\mathcal{G}%
}^{\left( 0\right) }}\left( {z},{z}\right) \;.
\end{gather*}

Similarly, for the two-particle Green's function
\begin{equation*}
\mathcal{G}_{II}\left( {{x}_{1}},{{x}_{2}},{{x}_{3}},{{x}_{4}}\right) =
\end{equation*}%
\begin{equation*}
-\frac{{{\left\langle {{T}_{\tau }}\left[ \psi \left( {{x}_{1}}\right) \psi
\left( {{x}_{2}}\right) \bar{\psi}\left( {x}_{{3}}\right) \bar{\psi}\left( {x%
}_{{4}}\right) \mathcal{S}\right] \right\rangle }_{0}}}{{{\left\langle
\mathcal{S}\right\rangle }_{0}}}\,,\,
\end{equation*}
we have
\begin{widetext}
\begin{gather*}
\mathcal{G}_{II}\left( {{x}_{1}},{{x}_{2}},{{x}_{3}},{{x}_{4}}\right) ={{\left\langle T\left[
\left( {{\psi \left( {{x}_{1}}\right) \bar{\psi}}}\left( {{x}_{3}}\right)
\right) \left( {{\psi \left( {x_{2}}\right) {\bar{\psi}}\left( {x_{4}}%
\right) }}\right) \right] \right\rangle }_{0}} \\
-\frac{1}{2{{\left\langle \mathcal{S}\right\rangle }_{0}}}\int
dz_{1}dz_{2}\mathcal{V}\left( z_{1}-z_{2}\right) {{\left\langle T\left[ \left( {{\psi
\left( {{x}_{1}}\right) \bar{\psi}}}\left( {{x}_{3}}\right) \right) \left( {{%
\psi \left( {x_{2}}\right) {\bar{\psi}}\left( {x_{4}}\right) }}\right)
\left( {{\psi \left( {z_{1}}\right) {\bar{\psi}}\left( {z_{1}}\right) }}%
\right) \left( {{\psi \left( {z}_{2}\right) {\bar{\psi}}}}\left(
z_{2}\right) \right) \right] \right\rangle }_{0}} \\
=\left\vert
\begin{array}{ll}
\Delta \left( x{_{1}},{{x}_{3}}\right)  & \Delta \left( {x_{1}},{x_{4}}%
\right)  \\
\Delta \left( x{_{2}},{{x}_{3}}\right)  & \Delta \left( x{_{2}},x{_{4}}%
\right)
\end{array}%
\right\vert -\frac{1}{2}\int dz_{1}dz_{2}\mathcal{V}\left( z_{1}-z_{2}\right)
\left\vert
\begin{array}{llll}
{\;\,0} & {\;\,0} & \Delta \left( x{_{1}},{z_{1}}\right)  & \Delta \left( {x_{1}},z{%
_{2}}\right)  \\
{\;\,0} & {\;\,0} & \Delta \left( x{_{2}},{z_{1}}\right)  & \Delta \left( {x_{2}},z{%
_{2}}\right)  \\
\;\Delta \left( z{_{1}},{{x}_{3}}\right)  & \;\Delta \left( z{_{1}},{x_{4}}%
\right)  & \Delta \left( z{_{1}},{z_{1}}\right)  & \Delta \left( z{_{1}},z{%
_{2}}\right)  \\
\;\Delta \left( z{_{2}},{{z}_{3}}\right)  & \;\Delta \left( z{_{2}},x{_{4}}%
\right)  & \Delta \left( z{_{2}},{z_{1}}\right)  & \Delta \left( z{_{2}},z{%
_{2}}\right)
\end{array}%
\right\vert  \\
={{\mathcal{G}}^{\left( 0\right) }}\left( {{x}_{1}},{{x}_{3}}\right) {{%
\mathcal{G}}^{\left( 0\right) }}\left( {{x}_{2}},{{x}_{4}}\right) -{{%
\mathcal{G}}^{\left( 0\right) }}\left( {{x}_{1}},{{x}_{4}}\right) {{\mathcal{%
G}}^{\left( 0\right) }}\left( {{x}_{2}},{x}_{{3}}\right)  \\
-\int dz_{1}dz_{2}{{\mathcal{G}}^{\left( 0\right) }}\left( {{x}_{1}},{z_{1}}%
\right) {{\mathcal{G}}^{\left( 0\right) }}\left( {{x}_{2}},{{z}_{2}}\right)
\mathcal{V}\left( z_{1}-z_{2}\right) \left[ {{\mathcal{G}}^{\left( 0\right) }}\left( {{%
z}_{1}},{{x}_{3}}\right) {{\mathcal{G}}^{\left( 0\right) }}\left( {{z}_{2}},{%
{x}_{4}}\right) -{{\mathcal{G}}^{\left( 0\right) }}\left( {{z}_{1}},{{x}_{4}}%
\right) {{\mathcal{G}}^{\left( 0\right) }}\left( {{z}_{2}},{x}_{{3}}\right) %
\right] \;.
\end{gather*}
\end{widetext}

\section{\label{conc}Conclusions}
The above theorem has an important consequence. In fact, it establishes a perfect coincidence between the vacuum expectation values of the chronological products of $n$  pairs of field operators and the  $n$--order determinant. If to present this determinant as the sum of the elements and cofactors of one any row or column, and thereafter to use again the indicated coincidence for the  $\left( n-1 \right)$--order determinants included in each summand, we will return to the generalized Wick's theorem. Alternatively, we can select in the our  $n$--order determinant arbitrary $m$  rows or columns ($1<m<n$) and use the {\it Generalized Laplace's Expansion}\/~\cite{Korn1961} for its presentation as the sum of the products of all  $m$--rowed minors using these rows (or columns) and their algebraic complements. Then, taking into account our theorem, we obtain a representation of the vacuum expectation values of the chronological products of $n$  pairs of field operators as the sum of the products of vacuum expectation values of the chronological products of $m$  pairs of operators and vacuum expectation values of the chronological products of $n-m$  pairs. The number of terms in this sum is equal to  $n!/m!\left(n-m\right)!$. This decomposition can be useful for the summation of blocks of diagrams.

Representations (\ref{fermi}) and (\ref{perm}) not only greatly simplify all calculations, but also allow one to perform them using a computer with programs of symbolic mathematics~\cite{Wolfram2003}.

\end{document}